\documentclass[aps,prl,twocolumn,groupedaddress,showpacs,floatfix]{revtex4}
\usepackage{dcolumn}
\usepackage{amsmath}
\usepackage{amssymb}
\usepackage{graphicx}
\usepackage{bm}
\usepackage[T1]{fontenc}
\usepackage{color}
\usepackage{url}
\usepackage{rotating}

\begin{document}

\title{Determining the critical coupling of explosive synchronization transitions in scale-free networks by mean-field approximations}

\author{Thomas Kau\^{e} Dal'Maso Peron}
\affiliation{Instituto de F\'{\i}sica
de S\~{a}o Carlos, Universidade de S\~{a}o Paulo, Av. Trabalhador
S\~{a}o Carlense 400, Caixa Postal 369, CEP 13560-970, S\~{a}o
Carlos, S\~ao Paulo, Brazil}
\author{Francisco A. Rodrigues}
\email{francisco@icmc.usp.br}
\affiliation{Departamento de Matem\'{a}tica Aplicada e Estat\'{i}stica, Instituto de Ci\^{e}ncias Matem\'{a}ticas e de Computa\c{c}\~{a}o,
Universidade de S\~{a}o Paulo, Caixa Postal 668,13560-970 S\~{a}o Carlos,  S\~ao Paulo, Brazil}

\begin{abstract}
Explosive synchronization can be observed in scale-free networks when Kuramoto oscillators have natural frequencies equal to their number of connections. In the current work, we took into account mean-field approximations to determine the critical coupling of such explosive synchronization. The obtained equation for the critical coupling has an inverse dependence with the network average degree. This expression differs from that calculated when the frequency distributions are unimodal and even. In this case, the critical coupling depends on the ratio between the first and second statistical moments of the degree distribution. We also conducted numerical simulations to verify our analytical results.
\end{abstract}


\maketitle

\section{Introduction}

Synchronization of coupled oscillators has been intensively studied because of its ubiquity in the real world~\cite{Barrat08:book, Arenas08:PR}. When a collection of oscillators are coupled as a network, it can be observed the emergence of a synchronous state~\cite{Pikovsky03, Arenas08:PR}. Such onset of coherent collective behavior has been verified between neurons in the central nervous system, communication networks, power grids, social interactions, animal behavior, ecosystems and circadian rhythm~\cite{Arenas08:PR}.

The level of synchronization of a system is the consequence of a combination of the type of oscillators, the connectivity organization, the time-delay and the interaction function~\cite{Barrat08:book, Arenas08:PR}. Particularly, the network topology has a strong influence on the value of the critical coupling~\cite{Moreno04:EPL,Arenas06:PRL, Zhou06:Chaos, Gomez07:PRL, Gomez07:PRL2} and on the stability of the fully synchronized state~\cite{Pecora98:PRL, Barahona02:PRL,Nishikawa03:PRL, Arenas08:PR}. For instance, Watts and Strogatz~\cite{Watts98:Nature} verified that the decrease in the average shortest path length in small-world networks facilitates a more efficient coupling and therefore enhances the level of synchronization. In addition, Nishikawa et al.~\cite{Nishikawa03:PRL} suggested that networks with an homogeneous degree distribution are more synchronizable than heterogenous ones.

The network structure is not only important to enhance the level of synchronization, but also to permit the occurrence of phase transitions. Indeed,  many works have verified second-order phase transitions in networks of Kuramoto oscillators~\cite{Arenas08:PR}.
Recently, Garde\~{n}es et al.~\cite{Gardenes011:PRL} showed that a first-order nonequilibrium synchronization transition can occur in scale-free networks. They suggested that this event is a consequence of a positive correlation between the heterogeneity of the connections and the natural frequencies of the oscillators~\cite{Gardenes011:PRL}. First-order phase transitions were also obtained experimentally and numerically by considering a network of R\"{o}sller units~\cite{Leyva012}. Indeed, such phenomena is attracting the interest of many complex networks researchers (\emph{e.g.}~\cite{Peron012, Chen012, Leyva012}).

Although the explosive synchronization has been observed in scale-free networks, the analytical expression that describes the critical coupling has not been determined yet. In the current work, we obtained such expression by considering mean-field approximations. We verified that the critical coupling has a inverse dependence with the network average degree. Our analytical results are compared with numerical simulations.

The Kuramoto model considers a set of $N$ oscillators coupled by the sine of their phase differences and phase oscillators at arbitrary frequencies~\cite{Acebron05:RMP}. Each oscillator is characterized by its phase $\theta_{i}(t)$, $i=1,...,N$. In complex networks, each oscillator $i$ obeys an equation of motion defined as
\begin{equation}\label{eq:kuramoto}
\frac{d\theta_{i}}{dt}= \omega_i + \lambda \sum_{i=1}^{N}A_{ij} \sin(\theta_j - \theta_i), \quad i=1,\ldots,N,
\end{equation}
where $\lambda$ is the coupling strength, $\omega_i$ is the natural frequency of oscillator $i$, and $A_{ij}$ are the elements of the adjacency matrix $A$, so that $A_{ij}=1$ when nodes $i$ and $j$ are connected while $A_{ij}=0$ otherwise. The general Kuramoto model considers a random distribution of the natural frequencies and phases according to a specific distribution $g(\omega)$~\cite{Barrat08:book, Arenas08:PR}. In most of the cases, the frequency distributions are unimodal and symmetric around a mean value $\omega_{0}$~\cite{Arenas08:PR}.

In the current work, we considered a modified version of the Kuramoto model as proposed by  Garde\~{n}es et. al.~\cite{Gardenes011:PRL}. More specifically, the natural frequency $\omega_{i}$ of the node $i$ was assigned to be equal to the node's degree $k_{i}$, \emph{i.e.}, $\omega_{i}=k_{i}$. Therefore, $g(\omega)=P(k)$, in which $P(k)$ is the degree distribution of the number of connections. This choice for the frequency distribution leads to the explosive synchronization in scale-free networks~\cite{Gardenes011:PRL}. Garde\~{n}es et. al. verified that this effect is due exclusively to the positive correlation between the network structure and the dynamics. When this correlation is broken, a first-order transition is no longer observed, whereas a second-order transition occurs~\cite{Gardenes011:PRL}.

In order to analyze the interplay between structure and dynamics in the Garde\~{n}es et. al. model, we considered the mean-field approach proposed by Ichinomiya~\cite{Ichinomiya04:PRE}. First, we characterized the network by its degree distribution $P(k)$ and introduced the density of the nodes with phase $\theta$ at time $t$  for a given degree $k$, denoted by $\rho(k;\theta,t)$, which is normalized according to
\begin{equation}
\int_{0}^{2\pi} \rho(k;\theta,t) d\theta=1.
\label{eq:normalization}
\end{equation}
The continuum limit of Eq.~\ref{eq:kuramoto} is taken by considering the absence of degree correlation between the nodes in the network. Observe that this is a typical assumption in mean-field approximation~\cite{Barrat08:book}. In this regime, the probability that a random edge is attached with a node with degree $k$ and phase $\theta$ at time $t$ is given as
\begin{equation}
\frac{ k  P(k)\rho(k;\theta,t) }{\left\langle k \right\rangle},
\label{eq:probability}
\end{equation}
where $\left\langle k \right\rangle$ is the network average degree. Replacing $\omega_{i}~=~k_{i}$ in Eq.~\ref{eq:kuramoto} and taking the continuum limit in the mean-field approach using Eq.~\ref{eq:probability}, we obtained
\begin{equation}
\frac{d\theta(t)}{dt} = k + \lambda k \int dk'\int d\theta' \frac{k'P(k')}{\left\langle k \right\rangle} \rho(k;\theta',t) \sin(\theta - \theta').
\label{eq:kuramoto_continuum}
\end{equation}
The order parameter, which quantifies the level synchronization of the network, is defined as~\cite{Ichinomiya04:PRE,Restrepo05:PRE}
\begin{equation}
r e^{i\psi(t)} = \frac{1}{\left\langle k \right\rangle}\int dk \int d\theta k P(k) \rho(k;\theta;t) e^{i\theta},
\label{eq:order}
\end{equation}
where $0 \leq r \leq 1 $ and $\psi(t)$ stands the average frequency of the oscillators. 

Multiplying Eq.\ref{eq:order} by $e^{-i\theta'}$, taking the imaginary part and including in Eq.~\ref{eq:kuramoto_continuum}, we obtained
\begin{equation}
\frac{d\theta}{dt} = k + \lambda k r\sin(\psi - \theta),
\label{eq:kuramoto_mean_field}
\end{equation}
which is the Eq.~\ref{eq:kuramoto_continuum} written in terms of the order parameter.

In order to let the equations of motion in function of known parameters of the network, we set a reference rotating frame $\psi(t)=\Omega t$, where $\Omega$ is the average frequency of the network. In the case of the Garde\~{n}es et al. model, \emph{i.e.}  $g(\omega)=P(k)$, the average frequency is equal to the network average degree  ($\Omega=\left\langle k \right\rangle$)~\cite{Gardenes011:PRL}. Defining a new variable as $\phi(t) \equiv \theta(t) - \psi(t)$ and replacing in   Eq.~\ref{eq:kuramoto_mean_field}, we obtained
\begin{equation}
\frac{d\phi}{dt} = (1 - \lambda r\sin\phi)k - \left\langle k \right\rangle.
\label{eq:kuramoto_mean_field_phi}
\end{equation}

We redefined the density of oscillators $\rho$ in terms of the new variable $\phi$, \emph{i.e.} $\rho = \rho(k;\phi,t)$. This density of oscillators must satisfy the continuity equation~\cite{Ichinomiya04:PRE}
\begin{equation}
\frac{\partial \rho(k;\phi,t)}{\partial t} + \frac{\partial }{\partial \phi}\{v_{\phi} \rho(k;\phi,t)  \}=0,
\label{eq:cont_eq}
\end{equation}
where $v_{\phi}= \frac{d\phi}{dt}$. Since we were interested in the analysis of the steady state of the system, we obtained the time-independent solutions of Eq.~\ref{eq:cont_eq}, \emph{i.e.}
\begin{equation}
\rho(k;\phi)=\begin{cases}
\delta\left(\phi-\arcsin\left[\frac{1}{\lambda r}\left(\frac{k-\left\langle k\right\rangle }{k}\right)\right]\right) & \mbox{ if }\frac{\left|k-\left\langle k\right\rangle \right|}{k}\leq \lambda r,\\
\frac{A(k)}{\left|\left(k-\left\langle k\right\rangle \right)-\lambda kr\sin\phi\right|} & \mbox{otherwise,}
\end{cases}
\label{eq:prob_solutions}
\end{equation}
where $\delta(\cdot)$ is the Dirac delta function and $A(k)$ is the normalization factor. The first solution is respective to the synchronous state, \emph{i.e.} $\frac{d\phi}{dt} = 0$, corresponding to the oscillators which are entrained by the mean field. On the other hand, the second one is the density of the non-entrained oscillators. \emph{i.e.} $\rho(k;\phi) \sim \frac{1}{| v_{\phi}|}$~\cite{Pikovsky03, Ichinomiya04:PRE}. Thus, to compute the integrals in Eq.~\ref{eq:order}, we redefined it in terms of the variable $\phi$ and separated the contribution of entrained and non-entrained oscillators
\begin{eqnarray}\nonumber
\left\langle k\right\rangle r & = & \int\left[\int_{\frac{\left|k-\left\langle k\right\rangle \right|}{k}\leq\lambda r}dk\right.\\
& + & \left.\int_{\frac{\left|k-\left\langle k\right\rangle \right|}{k}>\lambda r} dk\right]P(k)k\rho(k;\phi)e^{i\phi}d\phi.
\label{eq:r_II}
\end{eqnarray}

Rewriting the second integral in Eq.~\ref{eq:r_II} and noting that  $\rho(k;\phi)$ is $\pi$-periodic in $\phi$, we obtained
\begin{eqnarray*}
\int_{0}^{2\pi}\int_{\left\langle k\right\rangle /\left(1-\lambda r\right)}^{\infty}P(k)k\frac{\sqrt{(k-\left\langle k\right\rangle )^{2}-k^{2}\lambda^{2}r^{2}}}{2\pi\left(k-\left\langle k\right\rangle -k\lambda r\sin\phi\right)}dkd\phi\\
+\int_{0}^{2\pi}\int_{k_{\min}}^{\left\langle k\right\rangle /(1+\lambda r)}P(k)k\frac{\sqrt{(k-\left\langle k\right\rangle )^{2}-k^{2}\lambda^{2}r^{2}}}{2\pi\left(\left\langle k\right\rangle -k+k\lambda r\sin\phi\right)}dkd\phi & =0
\end{eqnarray*}
where $k_{\min}$ is the minimum degree in the network. Thus, only the contribution of the oscillators entrained in the mean-field is accounted in the summation of Eq.~\ref{eq:r_II}:
\begin{eqnarray}\nonumber
\left\langle k\right\rangle r & = & \int_{\left\langle k\right\rangle /\left(1+\lambda r\right)}^{\left\langle k\right\rangle /\left(1-\lambda r\right)}\exp\left[i\arcsin\left(\frac{1}{\lambda r}\left(\frac{k-\left\langle k\right\rangle }{k}\right)\right)\right]\\
& \times & kP(k)dk
\label{eq:r_III}
\end{eqnarray}

From the imaginary parte of Eq.~\ref{eq:r_III} we obtained
\begin{equation}
\int_{\left\langle k\right\rangle /\left(1+\lambda r\right)}^{\left\langle k\right\rangle /\left(1-\lambda r\right)}kP(k)\frac{1}{\lambda r}\left(\frac{k-\left\langle k\right\rangle }{k}\right)dk=0,
\label{eq:r_imaginary_part}
\end{equation}
and from the real part,
\begin{equation}
\left\langle k\right\rangle r=\int_{\left\langle k\right\rangle /\left(1+\lambda r\right)}^{\left\langle k\right\rangle /\left(1-\lambda r\right)}kP(k)\sqrt{1-\frac{1}{\lambda^{2}r^{2}}\left(\frac{k-\left\langle k\right\rangle }{k}\right)^{2}}dk.
\label{eq:r_real_part}
\end{equation}
Considering $x=(k-\left\langle k\right\rangle)/\lambda r$, we obtained
\begin{eqnarray}\nonumber
\left\langle k\right\rangle r & = & \lambda r\int_{-\left\langle k\right\rangle /\left(1+\lambda r\right)}^{\left\langle k\right\rangle /\left(1-\lambda r\right)}P(\lambda rx+\left\langle k\right\rangle )(\lambda rx+\left\langle k\right\rangle )\\
& \times & \sqrt{1-\left(\frac{x}{\lambda rx+\left\langle k\right\rangle }\right)^{2}}dx.
\label{eq:r_real_part_II}
\end{eqnarray}
For $r\neq 0$ and letting $r\rightarrow 0^{+}$,
\begin{equation}
\left\langle k\right\rangle =\lambda\int_{-\left\langle k\right\rangle }^{\left\langle k\right\rangle }P(\left\langle k\right\rangle )\left\langle k\right\rangle \sqrt{1-\left(\frac{x}{\left\langle k\right\rangle }\right)^{2}}dx,
\label{eq:r_>0}
\end{equation}
we reached to the critical coupling
\begin{equation}
\lambda_{c} = \frac{2}{\pi \left\langle k\right\rangle P\left( \left\langle k\right\rangle \right)}.
\label{eq:critical_coupling}
\end{equation}
Therefore, the critical coupling presents an inverse dependence with the average network degree and $ P\left( \left\langle k\right\rangle \right)$. This dependence is very different from that observed when it is taken into account other types of frequency distribution $g(\omega)$. For instance, if $g(\omega)$ is symmetric about a single local maximum $\omega_{0}$ (\emph{e.g.} $\omega_{0}=0$) the critical coupling is given as~\cite{Ichinomiya04:PRE,Restrepo05:PRE}
\begin{equation}
\lambda_{c}^{(0)} = \frac{2}{\pi g(0)} \frac{\left\langle k\right\rangle}{\left\langle k^{2}\right\rangle}.
\label{eq:critical_coupling_s}
\end{equation}
Thus, for scale-free networks, as $N \rightarrow \infty$ the critical coupling $\lambda_{c}^{(0)}$ become smaller, since the ratio $\left\langle k\right\rangle / \left\langle k^{2}\right\rangle $ diverges. On the other hand, for $g(\omega)=P(k)$, this effect should not be observed when $N\rightarrow \infty$, because the critical coupling depends only on the average degree $\left\langle k\right\rangle$ of the network.

\begin{figure}[!tpb]
\centerline{\includegraphics[width=1\linewidth]{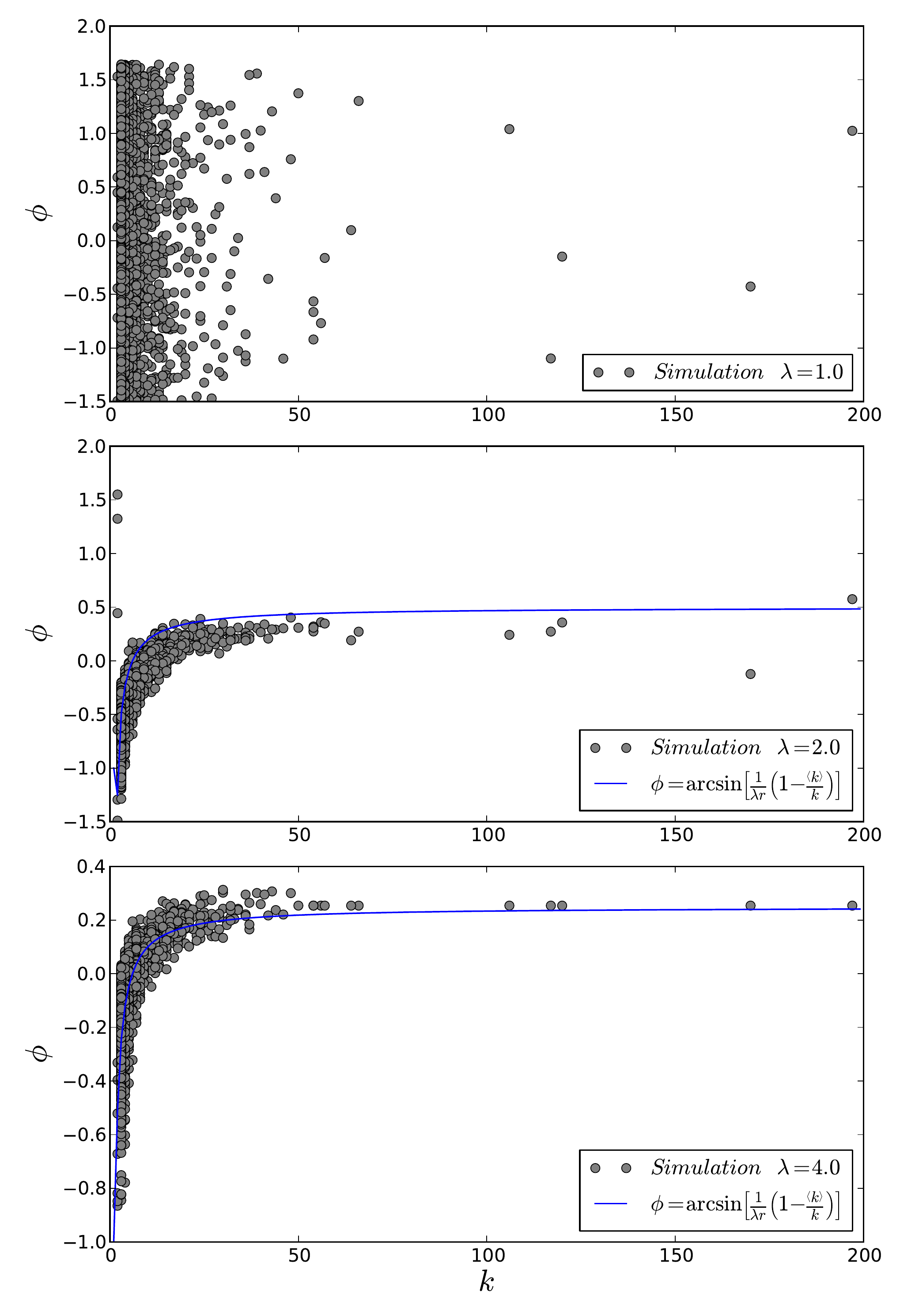}}
  \caption{Distribution of the phases $\phi$ as a function of the degree $k$ for a BA network with $N=3\cdot 10^{3}$ nodes and average degree $\left\langle k\right\rangle=6$.}
  \label{Fig:phase_dispersion}
\end{figure}

\begin{figure}[!tpb]
\centerline{\includegraphics[width=1\linewidth]{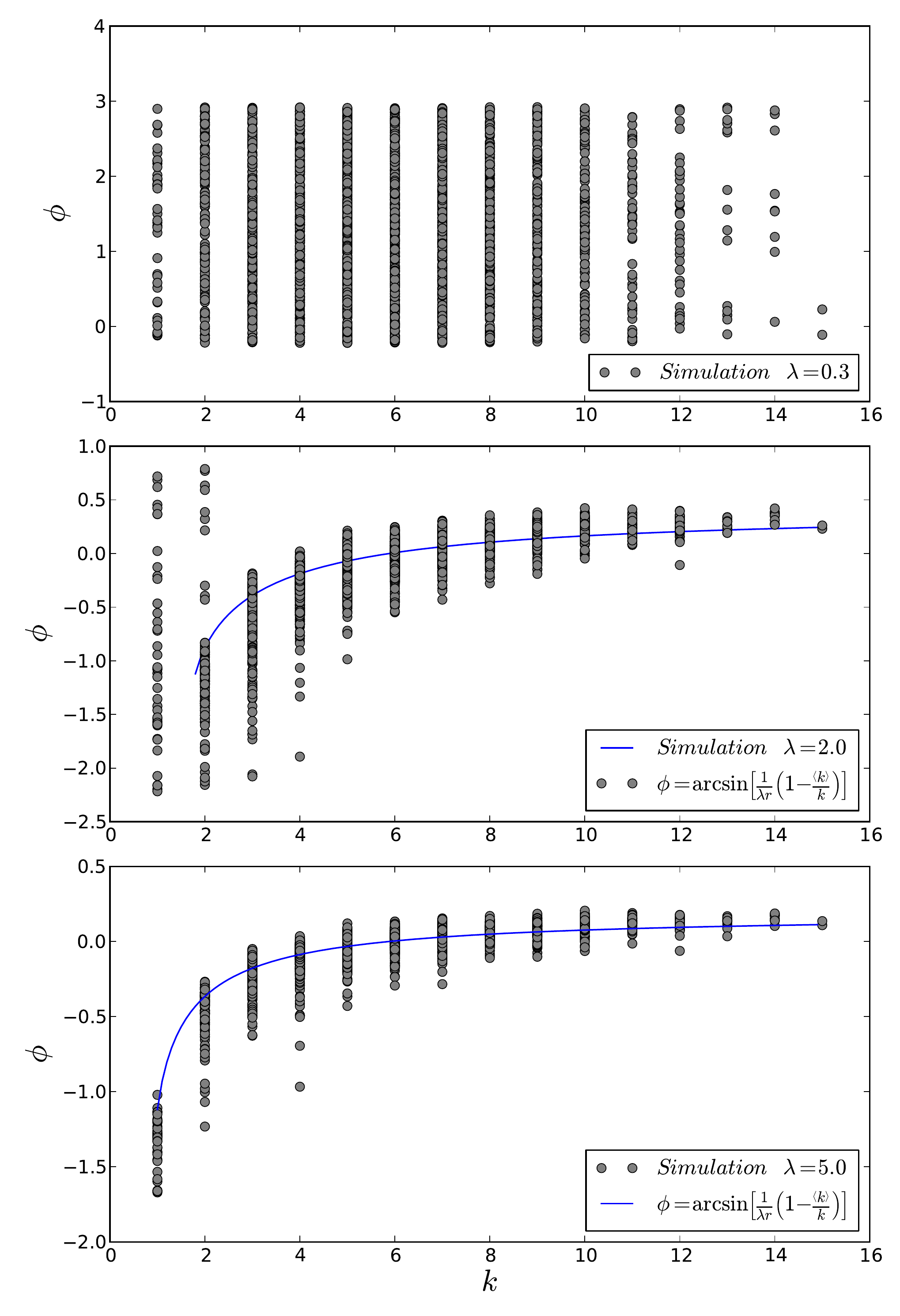}}
  \caption{Distribution of the phases $\phi$ as a function of the degree $k$ for a ER network with $N=3\cdot 10^{3}$ nodes and average degree $\left\langle k\right\rangle=6$ .}
  \label{Fig:phase_dispersion2}
\end{figure}

\begin{figure}[!tpb]
\centerline{\includegraphics[width=1\linewidth]{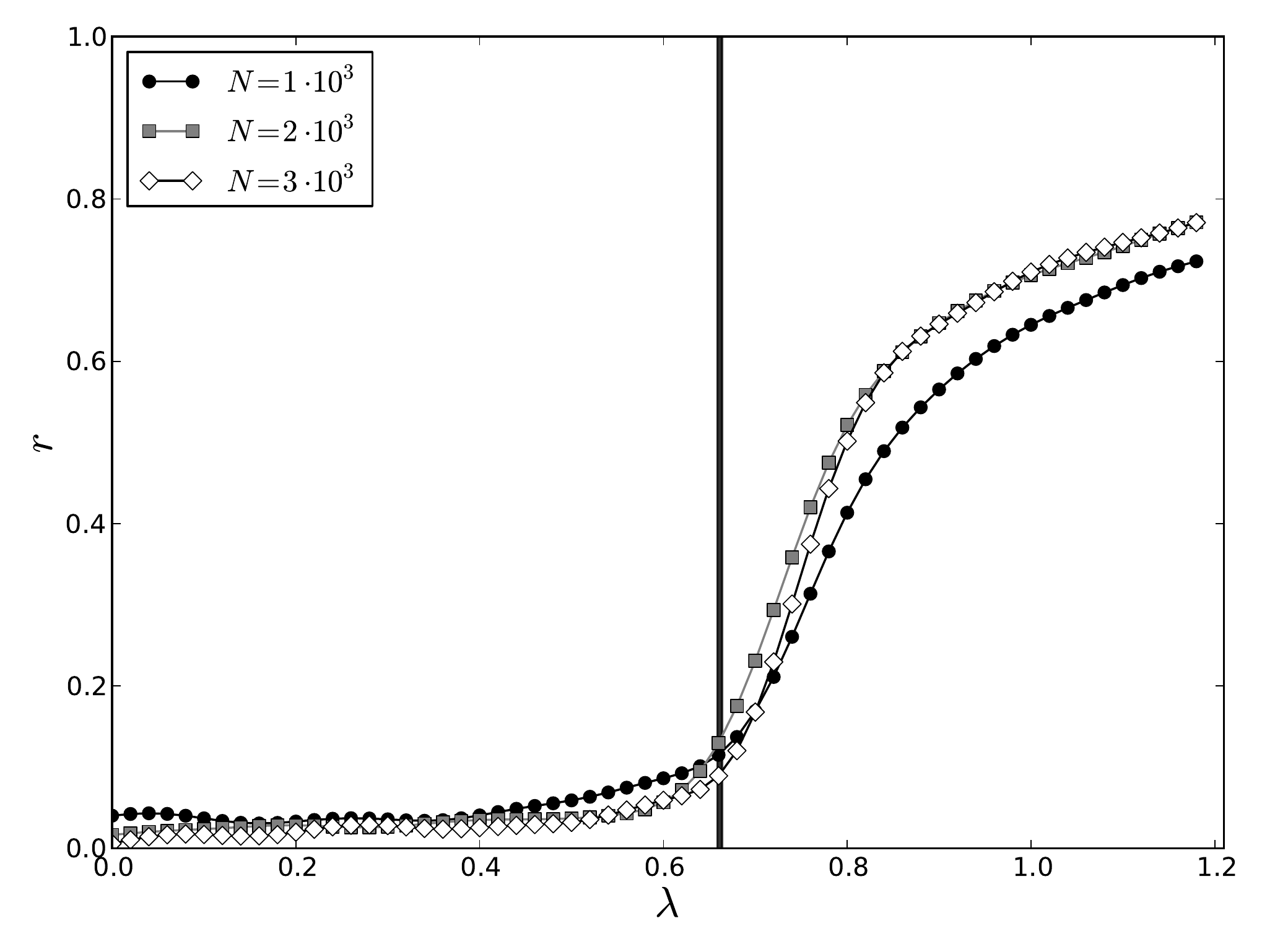}}
  \caption{Synchronization diagram for ER networks with forward continuation of the coupling strength $\lambda$ with steps of $\delta \lambda=0.02$. The networks have $N=1\cdot 10^{3},\mbox{ }2\cdot 10^{3}\mbox{ and }3\cdot 10^{3}$ and the same average degree $\left\langle k \right\rangle=6$. The black bar depicts
the theoretical value for the critical coupling $\lambda_c$.}
  \label{Fig:erdos}
\end{figure}

In order to check the validity of Eq.~\ref{eq:prob_solutions} and~\ref{eq:critical_coupling}, we considered numerical simulation. We increased the coupling strength $\lambda$ adiabatically and computed the stationary value of the global coherence $r$ for each value $\lambda_{0},\lambda_{0}+\delta \lambda,...,\lambda_{0}+n\delta\lambda$, with increments $\delta \lambda=0.02$, as done in~\cite{Gardenes011:PRL}.  Fig.~\ref{Fig:phase_dispersion}  shows the dispersion of the phases $\phi$ as function of the node's degree $k$ for a BA network with $N=10^{3}$ nodes with $\left\langle k \right\rangle=6$. As we can see in this figure, for $\lambda=2.0$ the system starts to present partial synchronization, suggesting that the critical is between $\lambda=1.0$ and $\lambda=2.0$. Note that for $\lambda=4.0$, the numerical results of the phases $\phi$ are in good agreement with the theoretical solution, specially for the highly connected nodes. Fig.~\ref{Fig:phase_dispersion2} also presents the dependence of the phases $\phi$ on the degree $k$ for a Erd\H{o}s-R\'{e}nyi (ER) network with $N=10 ^{3}$ and $\left\langle k \right\rangle=6$. As in Fig.~\ref{Fig:phase_dispersion}, we observed the same behavior for the ER network, as the coupling $\lambda$ becomes higher, the phases approaches the theoretical solution. Therefore, our results suggest that the solution of $\rho(k;\phi)$, given by Eq.~\ref{eq:prob_solutions}, is valid.

\begin{figure}[!tpb]
\centerline{\includegraphics[width=1\linewidth]{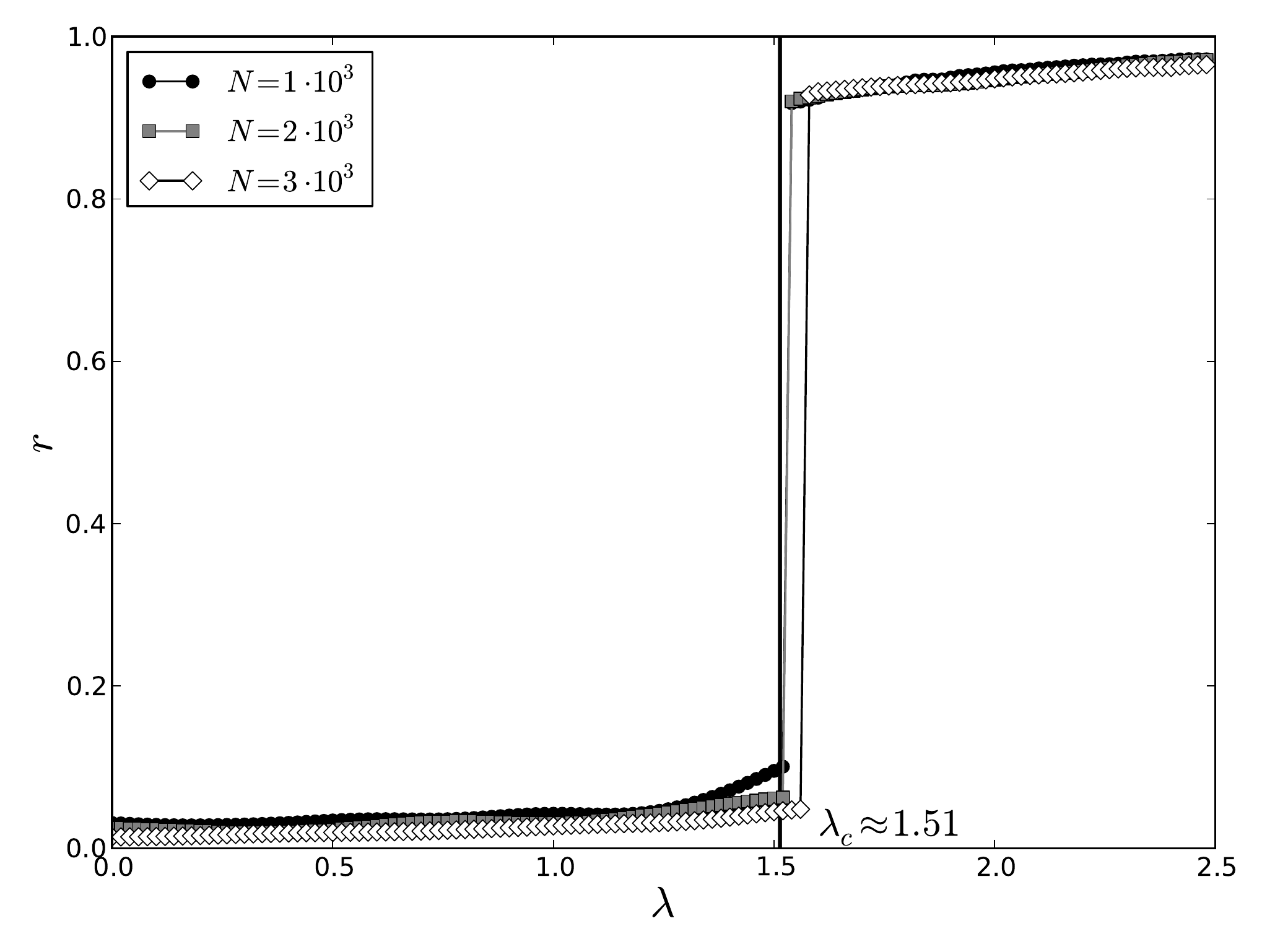}}
  \caption{Synchronization diagram for BA networks with forward continuation of the coupling strength $\lambda$ with steps of $\delta \lambda=0.02$. The networks have $N=1\cdot 10^{3},\mbox{ }2\cdot 10^{3}\mbox{ and }3\cdot 10^{3}$ and the same average degree $\left\langle k \right\rangle=6$. The black bar depicts the theoretical value for the critical coupling $\lambda_c$.}
  \label{Fig:barabasi}
\end{figure}

Once we have verified the validity of Eq.~\ref{eq:prob_solutions} , we estimated the critical coupling considering numerical data. We took into account an ensemble of $N_{net}$ networks, $\left\{N_{1},N_{2},...,N_{net}\right\}$, with the same number of nodes $N$ and same average degree $\left\langle k \right\rangle=6$. We estimated the critical coupling $\lambda_{c}$ as an average over this ensemble by using Eq.~\ref{eq:critical_coupling}. In this way, we obtained $\lambda_{c}^{(N=1000)} \cong 0.65$, $\lambda_{c}^{(N=2000)} \cong 0.65$ and $\lambda_{c}^{(N=3000)} \cong 0.66$ for the ER networks. 
Fig.~\ref{Fig:erdos} shows the coherence diagram of $r$ as function of $\lambda$ for ER networks with $N= 10^{3}$, $2 \cdot 10^{3}$ and $3 \cdot 10^{3}$ nodes. As we can see in this figure, the critical coupling $\lambda_{c}$ does not depend extensively on the total number of nodes $N$ in the network, since the theoretical estimation for the critical coupling in Eq.~\ref{eq:critical_coupling} depends only on the network average degree.

\begin{figure}[!tpb]
\centerline{\includegraphics[width=1\linewidth]{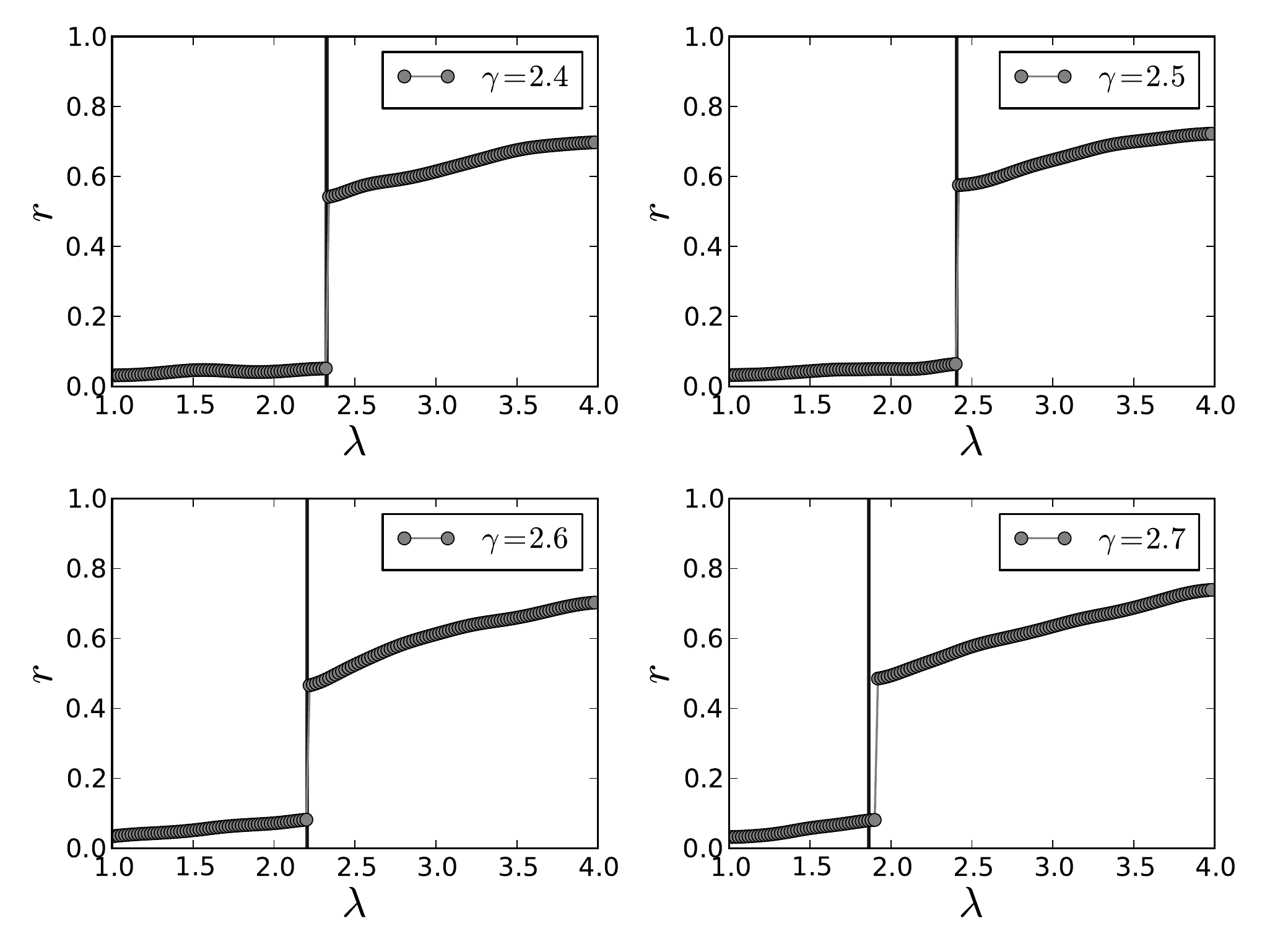}}
  \caption{Coherence diagrams for configurational models with degree distribution $P(k) \sim k^{-\gamma}$: (a) $\gamma = 2.4$ (b) $\gamma = 2.5$ (c) $\gamma = 2.6$ and (d) $\gamma = 2.7$, with forward continuation of the coupling strength $\lambda$ with steps of $\delta \lambda = 0.02$.}
  \label{Fig:configurational}
\end{figure}

In order to conducted numerical simulations to verify the validity of Eq.~\ref{eq:critical_coupling},  we considered the model by Barab\'{a}si and Albert (BA) and a configuration model. Networks generated by the BA model are characterized by a distribution of connections following a power law, \emph{i.e.} $P(k) \sim k^{-3}$~\cite{Barabasi99:Science}. The configuration model allows to generate networks with a given degree sequence~\cite{Newman010:book}. Fig.~\ref{Fig:barabasi} shows the coherence diagram for a BA network. Performing the same procedure described above to estimated the critical coupling, we obtained the values $\lambda_{c}^{(N=1000)} \cong 1.51$, $\lambda_{c}^{(N=2000)} \cong 1.51$ and $\lambda_{c}^{(N=3000)} \cong 1.52$, which are in good agreement with the results from the numerical simulation. Also,  Fig.~\ref{Fig:configurational} shows the explosive synchronization in scale-free networks with degree distribution $P(k) \sim k^{-\gamma}$ constructed using the configurational model~\cite{Newman010:book} with $\gamma=$ $2.4$, $2.5$, $2.6$, and $2.7$, considering degrees in the range $1\leq k \leq 100$.

When the frequency distribution, $g(\omega)$, is unimodal and even, the critical coupling tends to vanish as $N\rightarrow\infty$. On the other hand, the assumption that the natural frequencies are equal to the degree implies that the critical coupling does not suffer significant variations. This fact can be observed in Fig.~\ref{Fig:barabasi}. In addition, we did not verify that for the forward continuations of $\lambda$, the critical coupling increases extensively with the number of nodes. This result was obtained  in~\cite{Gardenes011:PRL}, where authors considered a star network as an approximation of scale-free networks. Therefore, although star networks exhibit the first order phase transition, the critical coupling does not have the same behavior as verified in scale-free networks.

The analysis performed in the current work helps to understand the relationship between the structure and the explosive synchronization in scale-free networks. The obtained expression for the critical coupling does not depends on the ratio $\left\langle k\right\rangle / \left\langle k^{2}\right\rangle $, as observed in the case when $g(\omega)$ is symmetric. Indeed, the obtained critical coupling has a inverse dependence with the network average degree, $\left\langle k\right\rangle$,  and $P(\left\langle k\right\rangle)$.

Francisco A. Rodrigues would like to acknowledge CNPq (305940/2010-4) and FAPESP (2010/19440-2) for the financial support given to this research. Thomas K. D. M. Peron would like to acknowledge Fapesp for the sponsorship provided. 

\bibliographystyle{apsrev}
\bibliography{paper}

\begin{thebibliography}{21}
\expandafter\ifx\csname natexlab\endcsname\relax\def\natexlab#1{#1}\fi
\expandafter\ifx\csname bibnamefont\endcsname\relax
  \def\bibnamefont#1{#1}\fi
\expandafter\ifx\csname bibfnamefont\endcsname\relax
  \def\bibfnamefont#1{#1}\fi
\expandafter\ifx\csname citenamefont\endcsname\relax
  \def\citenamefont#1{#1}\fi
\expandafter\ifx\csname url\endcsname\relax
  \def\url#1{\texttt{#1}}\fi
\expandafter\ifx\csname urlprefix\endcsname\relax\def\urlprefix{URL }\fi
\providecommand{\bibinfo}[2]{#2}
\providecommand{\eprint}[2][]{\url{#2}}

\bibitem[{\citenamefont{Barrat et~al.}(2008)\citenamefont{Barrat, Barthlemy,
  and Vespignani}}]{Barrat08:book}
\bibinfo{author}{\bibfnamefont{A.}~\bibnamefont{Barrat}},
  \bibinfo{author}{\bibfnamefont{M.}~\bibnamefont{Barthlemy}},
  \bibnamefont{and}
  \bibinfo{author}{\bibfnamefont{A.}~\bibnamefont{Vespignani}},
  \emph{\bibinfo{title}{Dynamical processes on complex networks}}
  (\bibinfo{publisher}{Cambridge University Press}, \bibinfo{year}{2008}).

\bibitem[{\citenamefont{Arenas et~al.}(2008)\citenamefont{Arenas,
  D{\'\i}az-Guilera, Kurths, Moreno, and Zhou}}]{Arenas08:PR}
\bibinfo{author}{\bibfnamefont{A.}~\bibnamefont{Arenas}},
  \bibinfo{author}{\bibfnamefont{A.}~\bibnamefont{D{\'\i}az-Guilera}},
  \bibinfo{author}{\bibfnamefont{J.}~\bibnamefont{Kurths}},
  \bibinfo{author}{\bibfnamefont{Y.}~\bibnamefont{Moreno}}, \bibnamefont{and}
  \bibinfo{author}{\bibfnamefont{C.}~\bibnamefont{Zhou}},
  \bibinfo{journal}{Physics Reports} \textbf{\bibinfo{volume}{469}},
  \bibinfo{pages}{93} (\bibinfo{year}{2008}).

\bibitem[{\citenamefont{Pikovsky et~al.}(2003)\citenamefont{Pikovsky,
  Rosenblum, and Kurths}}]{Pikovsky03}
\bibinfo{author}{\bibfnamefont{A.}~\bibnamefont{Pikovsky}},
  \bibinfo{author}{\bibfnamefont{M.}~\bibnamefont{Rosenblum}},
  \bibnamefont{and} \bibinfo{author}{\bibfnamefont{J.}~\bibnamefont{Kurths}},
  \emph{\bibinfo{title}{Synchronization: A universal concept in nonlinear
  sciences}}, vol.~\bibinfo{volume}{12} (\bibinfo{publisher}{Cambridge
  University Press}, \bibinfo{year}{2003}).

\bibitem[{\citenamefont{Moreno and Pacheco}(2004)}]{Moreno04:EPL}
\bibinfo{author}{\bibfnamefont{Y.}~\bibnamefont{Moreno}} \bibnamefont{and}
  \bibinfo{author}{\bibfnamefont{A.~F.} \bibnamefont{Pacheco}},
  \bibinfo{journal}{EPL (Europhysics Letters)} \textbf{\bibinfo{volume}{68}},
  \bibinfo{pages}{603} (\bibinfo{year}{2004}).

\bibitem[{\citenamefont{Arenas et~al.}(2006)\citenamefont{Arenas, Diaz-Guilera,
  and P{\'e}rez-Vicente}}]{Arenas06:PRL}
\bibinfo{author}{\bibfnamefont{A.}~\bibnamefont{Arenas}},
  \bibinfo{author}{\bibfnamefont{A.}~\bibnamefont{Diaz-Guilera}},
  \bibnamefont{and}
  \bibinfo{author}{\bibfnamefont{C.}~\bibnamefont{P{\'e}rez-Vicente}},
  \bibinfo{journal}{{Physical Review Letters}} \textbf{\bibinfo{volume}{96}},
  \bibinfo{pages}{114102} (\bibinfo{year}{2006}).

\bibitem[{\citenamefont{Zhou and Kurths}(2006)}]{Zhou06:Chaos}
\bibinfo{author}{\bibfnamefont{C.}~\bibnamefont{Zhou}} \bibnamefont{and}
  \bibinfo{author}{\bibfnamefont{J.}~\bibnamefont{Kurths}},
  \bibinfo{journal}{Chaos: An Interdisciplinary Journal of Nonlinear Science}
  \textbf{\bibinfo{volume}{16}}, \bibinfo{pages}{015104}
  (\bibinfo{year}{2006}).

\bibitem[{\citenamefont{G{\'o}mez-Gardenes
  et~al.}(2007)\citenamefont{G{\'o}mez-Gardenes, Moreno, and
  Arenas}}]{Gomez07:PRL}
\bibinfo{author}{\bibfnamefont{J.}~\bibnamefont{G{\'o}mez-Gardenes}},
  \bibinfo{author}{\bibfnamefont{Y.}~\bibnamefont{Moreno}}, \bibnamefont{and}
  \bibinfo{author}{\bibfnamefont{A.}~\bibnamefont{Arenas}},
  \bibinfo{journal}{Physical Review Letters} \textbf{\bibinfo{volume}{98}},
  \bibinfo{pages}{34101} (\bibinfo{year}{2007}).

\bibitem[{\citenamefont{G{\'o}mez-Garde{\~n}es
  et~al.}(2007)\citenamefont{G{\'o}mez-Garde{\~n}es, Moreno, and
  Arenas}}]{Gomez07:PRL2}
\bibinfo{author}{\bibfnamefont{J.}~\bibnamefont{G{\'o}mez-Garde{\~n}es}},
  \bibinfo{author}{\bibfnamefont{Y.}~\bibnamefont{Moreno}}, \bibnamefont{and}
  \bibinfo{author}{\bibfnamefont{A.}~\bibnamefont{Arenas}},
  \bibinfo{journal}{Physical Review E} \textbf{\bibinfo{volume}{75}},
  \bibinfo{pages}{066106} (\bibinfo{year}{2007}).

\bibitem[{\citenamefont{Pecora and Carroll}(1998)}]{Pecora98:PRL}
\bibinfo{author}{\bibfnamefont{L.~M.} \bibnamefont{Pecora}} \bibnamefont{and}
  \bibinfo{author}{\bibfnamefont{T.~L.} \bibnamefont{Carroll}},
  \bibinfo{journal}{Physical Review Letters} \textbf{\bibinfo{volume}{80}},
  \bibinfo{pages}{2109} (\bibinfo{year}{1998}).

\bibitem[{\citenamefont{Barahona and Pecora}(2002)}]{Barahona02:PRL}
\bibinfo{author}{\bibfnamefont{M.}~\bibnamefont{Barahona}} \bibnamefont{and}
  \bibinfo{author}{\bibfnamefont{L.}~\bibnamefont{Pecora}},
  \bibinfo{journal}{Physical Review Letters} \textbf{\bibinfo{volume}{89}},
  \bibinfo{pages}{54101} (\bibinfo{year}{2002}).

\bibitem[{\citenamefont{Nishikawa et~al.}(2003)\citenamefont{Nishikawa, Motter,
  Lai, and Hoppensteadt}}]{Nishikawa03:PRL}
\bibinfo{author}{\bibfnamefont{T.}~\bibnamefont{Nishikawa}},
  \bibinfo{author}{\bibfnamefont{A.}~\bibnamefont{Motter}},
  \bibinfo{author}{\bibfnamefont{Y.}~\bibnamefont{Lai}}, \bibnamefont{and}
  \bibinfo{author}{\bibfnamefont{F.}~\bibnamefont{Hoppensteadt}},
  \bibinfo{journal}{Physical Review Letters} \textbf{\bibinfo{volume}{91}},
  \bibinfo{pages}{14101} (\bibinfo{year}{2003}).

\bibitem[{\citenamefont{Watts and Strogatz}(1998)}]{Watts98:Nature}
\bibinfo{author}{\bibfnamefont{D.}~\bibnamefont{Watts}} \bibnamefont{and}
  \bibinfo{author}{\bibfnamefont{S.}~\bibnamefont{Strogatz}},
  \bibinfo{journal}{Nature} \textbf{\bibinfo{volume}{393}},
  \bibinfo{pages}{440} (\bibinfo{year}{1998}).

\bibitem[{\citenamefont{Gomez-Garde\~{n}es
  et~al.}(2011)\citenamefont{Gomez-Garde\~{n}es, Gomez, Arenas, and
  Moreno}}]{Gardenes011:PRL}
\bibinfo{author}{\bibfnamefont{J.}~\bibnamefont{Gomez-Garde\~{n}es}},
  \bibinfo{author}{\bibfnamefont{S.}~\bibnamefont{Gomez}},
  \bibinfo{author}{\bibfnamefont{A.}~\bibnamefont{Arenas}}, \bibnamefont{and}
  \bibinfo{author}{\bibfnamefont{Y.}~\bibnamefont{Moreno}},
  \bibinfo{journal}{Physical Review Letters} \textbf{\bibinfo{volume}{106}},
  \bibinfo{pages}{128701} (\bibinfo{year}{2011}).

\bibitem[{\citenamefont{Leyva et~al.}(2012)\citenamefont{Leyva,
  Sevilla-Escoboza, Buld{\'u}, Sendi{\~n}a-Nadal, G{\'o}mez-Garde{\~n}es,
  Arenas, Moreno, G{\'o}mez, Jaimes-Re{\'a}tegui, and Boccaletti}}]{Leyva012}
\bibinfo{author}{\bibfnamefont{I.}~\bibnamefont{Leyva}},
  \bibinfo{author}{\bibfnamefont{R.}~\bibnamefont{Sevilla-Escoboza}},
  \bibinfo{author}{\bibfnamefont{J.}~\bibnamefont{Buld{\'u}}},
  \bibinfo{author}{\bibfnamefont{I.}~\bibnamefont{Sendi{\~n}a-Nadal}},
  \bibinfo{author}{\bibfnamefont{J.}~\bibnamefont{G{\'o}mez-Garde{\~n}es}},
  \bibinfo{author}{\bibfnamefont{A.}~\bibnamefont{Arenas}},
  \bibinfo{author}{\bibfnamefont{Y.}~\bibnamefont{Moreno}},
  \bibinfo{author}{\bibfnamefont{S.}~\bibnamefont{G{\'o}mez}},
  \bibinfo{author}{\bibfnamefont{R.}~\bibnamefont{Jaimes-Re{\'a}tegui}},
  \bibnamefont{and}
  \bibinfo{author}{\bibfnamefont{S.}~\bibnamefont{Boccaletti}},
  \bibinfo{journal}{Physical Review Letters} \textbf{\bibinfo{volume}{108}}
  (\bibinfo{year}{2012}).

\bibitem[{\citenamefont{Peron and Rodrigues}(2012)}]{Peron012}
\bibinfo{author}{\bibfnamefont{T.~K. D.~M.} \bibnamefont{Peron}}
  \bibnamefont{and} \bibinfo{author}{\bibfnamefont{F.~A.}
  \bibnamefont{Rodrigues}}, \bibinfo{journal}{Arxiv preprint arXiv:1110.5377}
  (\bibinfo{year}{2012}).

\bibitem[{\citenamefont{Chen et~al.}(2012)\citenamefont{Chen, Huang, Shen, and
  Hou}}]{Chen012}
\bibinfo{author}{\bibfnamefont{H.}~\bibnamefont{Chen}},
  \bibinfo{author}{\bibfnamefont{F.}~\bibnamefont{Huang}},
  \bibinfo{author}{\bibfnamefont{C.}~\bibnamefont{Shen}}, \bibnamefont{and}
  \bibinfo{author}{\bibfnamefont{Z.}~\bibnamefont{Hou}},
  \bibinfo{journal}{Arxiv preprint arXiv:1204.1816}  (\bibinfo{year}{2012}).

\bibitem[{\citenamefont{Acebr{\'o}n et~al.}(2005)\citenamefont{Acebr{\'o}n,
  Bonilla, Vicente, Ritort, and Spigler}}]{Acebron05:RMP}
\bibinfo{author}{\bibfnamefont{J.~A.} \bibnamefont{Acebr{\'o}n}},
  \bibinfo{author}{\bibfnamefont{L.~L.} \bibnamefont{Bonilla}},
  \bibinfo{author}{\bibfnamefont{C.~J.~P.} \bibnamefont{Vicente}},
  \bibinfo{author}{\bibfnamefont{F.}~\bibnamefont{Ritort}}, \bibnamefont{and}
  \bibinfo{author}{\bibfnamefont{R.}~\bibnamefont{Spigler}},
  \bibinfo{journal}{Reviews of Modern Physics} \textbf{\bibinfo{volume}{77}},
  \bibinfo{pages}{137} (\bibinfo{year}{2005}).

\bibitem[{\citenamefont{Ichinomiya}(2004)}]{Ichinomiya04:PRE}
\bibinfo{author}{\bibfnamefont{T.}~\bibnamefont{Ichinomiya}},
  \bibinfo{journal}{Physical Review E} \textbf{\bibinfo{volume}{70}},
  \bibinfo{pages}{026116} (\bibinfo{year}{2004}).

\bibitem[{\citenamefont{Restrepo et~al.}(2005)\citenamefont{Restrepo, Ott, and
  Hunt}}]{Restrepo05:PRE}
\bibinfo{author}{\bibfnamefont{J.~G.} \bibnamefont{Restrepo}},
  \bibinfo{author}{\bibfnamefont{E.}~\bibnamefont{Ott}}, \bibnamefont{and}
  \bibinfo{author}{\bibfnamefont{B.~R.} \bibnamefont{Hunt}},
  \bibinfo{journal}{Physical Review E} \textbf{\bibinfo{volume}{71}},
  \bibinfo{pages}{036151} (\bibinfo{year}{2005}).

\bibitem[{\citenamefont{Barab{\'a}si and Albert}(1999)}]{Barabasi99:Science}
\bibinfo{author}{\bibfnamefont{A.-L.} \bibnamefont{Barab{\'a}si}}
  \bibnamefont{and} \bibinfo{author}{\bibfnamefont{R.}~\bibnamefont{Albert}},
  \bibinfo{journal}{Science} \textbf{\bibinfo{volume}{286}},
  \bibinfo{pages}{509} (\bibinfo{year}{1999}).

\bibitem[{\citenamefont{Newman}(2010)}]{Newman010:book}
\bibinfo{author}{\bibfnamefont{M.}~\bibnamefont{Newman}},
  \emph{\bibinfo{title}{Networks: an introduction}} (\bibinfo{publisher}{Oxford
  University Press}, \bibinfo{year}{2010}).

\end{thebibliography}

\end{document}